%% file: main.tex
\documentclass[10pt, conference, letterpaper]{IEEEtran}
\IEEEoverridecommandlockouts
\usepackage{cite}
\usepackage{amsmath,amssymb,amsfonts}
\usepackage{graphicx}
\usepackage{textcomp}
\usepackage{xcolor}
\usepackage{algpseudocode}
\usepackage[linesnumbered,ruled,noline]{algorithm2e}

\SetCommentSty{mycommfont}
\SetKwInput{KwInput}{Input}                
\SetKwInput{KwOutput}{Output}              
\SetKwBlock{Loop}{Loop}{}
\SetKwBlock{Func}{Function}{}
\SetKwBlock{While}{While}{}

\def\BibTeX{{\rm B\kern-.05em{\sc i\kern-.025em b}\kern-.08em
    T\kern-.1667em\lower.7ex\hbox{E}\kern-.125emX}}

\AtBeginDocument{%
  \providecommand\BibTeX{{%
    \normalfont B\kern-0.5em{\scshape i\kern-0.25em b}\kern-0.8em\TeX}}}

\usepackage{listings}
\newcommand{\ours}{\mbox{DynamicFL}\xspace}
\newcommand{\sota}{\textsf{Oort}\xspace}

\newcommand{\yogi}{\yogi{YoGi}\xspace}

\usepackage[english]{babel}
\usepackage{blindtext}
\usepackage{dsfont}

\usepackage{enumitem,kantlipsum}
\usepackage{multirow}
\usepackage{subfigure}
\usepackage{colortbl}

\usepackage{pifont}

\usepackage{array}
\newcolumntype{C}[1]{>{\centering\arraybackslash}p{#1}}

\usepackage{tikz}

\definecolor{green}{HTML}{3049D4}
\definecolor{codegreen}{rgb}{0,0.6,0}
\definecolor{codegray}{rgb}{0.5,0.5,0.5}
\definecolor{codepurple}{rgb}{0.58,0,0.82}
\definecolor{backcolour}{rgb}{0.95,0.95,0.92}

\lstdefinestyle{mystyle}{
    commentstyle=\color{codegreen},
    keywordstyle=\color{magenta},
    numberstyle=\tiny\color{codegray},
    stringstyle=\color{codepurple},
    basicstyle=\ttfamily\footnotesize,
    breakatwhitespace=false,         
    breaklines=true,                 
    captionpos=b,                    
    keepspaces=true,                 
    numbers=left,                    
    numbersep=10pt,                  
    showspaces=false,                
    showstringspaces=false,
    showtabs=false,                  
    tabsize=2
}

\begin{document}

\title{\ours: Balancing Communication Dynamics and Client Manipulation for Federated Learning \\
}


\author{\IEEEauthorblockN{
Bocheng Chen,
Nikolay Ivanov,   
Guangjing Wang,
Qiben Yan   
}                                     
\IEEEauthorblockA{
 Computer Science \& Engineering, Michigan State University, East Lansing, USA. \\
Email:\{chenboc1, ivanovn1, wanggu22, qyan\}@msu.edu
}}

\maketitle

\input{0_abstract}


\section{Introduction}
\label{sec:introduction}
\input{1_introduction}


\section{Background and Motivation}
\label{sec:background}
\input{2_background}


\section{\ours Framework}
\label{sec:4}
\input{4_design-detail}

\section{Evaluation}
\label{sec:evaluation}
\input{5_evaluation}


\section{Related Work}
\label{sec:related}
\input{6_relatedwork}


\section{Conclusion}
\label{sec:conclusion}
\input{7_conclusion}


\clearpage

\bibliographystyle{IEEEtran}
\bibliography{bibliography}


\end{document}

%% file: 0_abstract.tex
\begin{abstract}
Federated Learning (FL) is a distributed machine learning (ML) paradigm, aiming to train a global model by exploiting the decentralized data across millions of edge devices. Compared with centralized learning, FL preserves the clients' privacy by refraining from explicitly downloading their data. However, given the geo-distributed edge devices (e.g., mobile, car, train, or subway) with highly dynamic networks in the wild, aggregating all the model updates from those participating devices will result in inevitable long-tail delays in FL. This will significantly degrade the efficiency of the training process. To resolve the high system heterogeneity in time-sensitive FL scenarios, we propose a novel FL framework, DynamicFL, by considering the communication dynamics and data quality across massive edge devices with a specially designed client manipulation strategy. \ours  actively selects clients for model updating based on the network prediction from its dynamic network conditions and the quality of its training data. Additionally, our long-term greedy strategy in client selection 
tackles the problem of system performance degradation caused by short-term scheduling in a dynamic network. 
Lastly, to balance the trade-off between client performance evaluation and client manipulation granularity, we dynamically adjust the length of the observation window
in the training process to optimize the long-term system efficiency.  
Compared with the state-of-the-art client selection scheme in FL, \ours can achieve a better model accuracy while consuming only 18.9\% -- 84.0\% of the wall-clock time. Our component-wise and sensitivity studies further demonstrate the robustness of \ours under various real-life scenarios. 
\end{abstract}

%% file: 1_introduction.tex
Federated Learning (FL) enables globally distributed devices to participate in the model training without leaking raw data~\cite{mcmahan2017communication_fedavg,wang2023federated}. 
FL has been applied across various fields where millions of edge devices participate in the training with their local data, including connected autonomous vehicle's motion planning~\cite{fu2022selective,liang2022federated,donevski2021addressing}, activity prediction\cite{ouyang2021clusterfl,tu2021feddl,WiSIA_Sensys20}, text prediction\cite{leroy2019federated,chen2023toxicbot}, and smart home automation\cite{zhao2020privacy,wang2023federated2,wang2023graph,alhanahnah2022iotcom}.

\noindent
\textbf{Status Quo and Limitations.}
In the FL training process, a round of training comprises two steps: first, the server dispatches the model to each client; second, the server aggregates the uploaded model updates once all clients have completed their training. It can take hundreds of rounds for the model to achieve its final accuracy, which can take hours. One key problem in FL is achieving the desired model accuracy within a short time, commonly referred to as \textbf{time-to-accuracy}~\cite{lai_oort_2021,PyramidFL_MobiCom22}. On the server side, reducing the time required to achieve the final model accuracy can minimize the overall energy cost and bandwidth consumption for each client. On the client side, each client can receive the best model in less time. Time-to-accuracy is particularly important in time-sensitive FL tasks, such as collision avoidance for autonomous driving~\cite{fu2022selective,liang2022federated}, where model updates from vehicles should be gathered in real-time.  
In recent years, as the FL system relies on the participation of a large number of clients, there has been increasing attention on improving the efficiency of FL systems.
The client selection method~\cite{lai_oort_2021,PyramidFL_MobiCom22} is the state-of-the-art (SOTA) method for improving the efficiency
of the FL system. This method has shown significant improvement in the time-to-accuracy of FL training.

However, in a realistic network scenario,  clients often experience varying and fluctuating network conditions. This presents a significant challenge for FL systems. For example, when a client selected by the server happens to be in a tunnel, the downgraded network connectivity creates a bottleneck in the training round and results in a significant delay during the model upload stage. Moreover, for clients prioritized by the selection algorithm due to their fast communication speed in previous rounds, an unstable network can cause adverse outcomes when they are faced with unstable networks.

\begin{figure}[t]
\centering 
\subfigure[Dynamic bandwidth affects the time-to-accuracy (`w' indicates with dynamic bandwidth, `w/o' indicates without it).]{\includegraphics[width=0.42\linewidth]{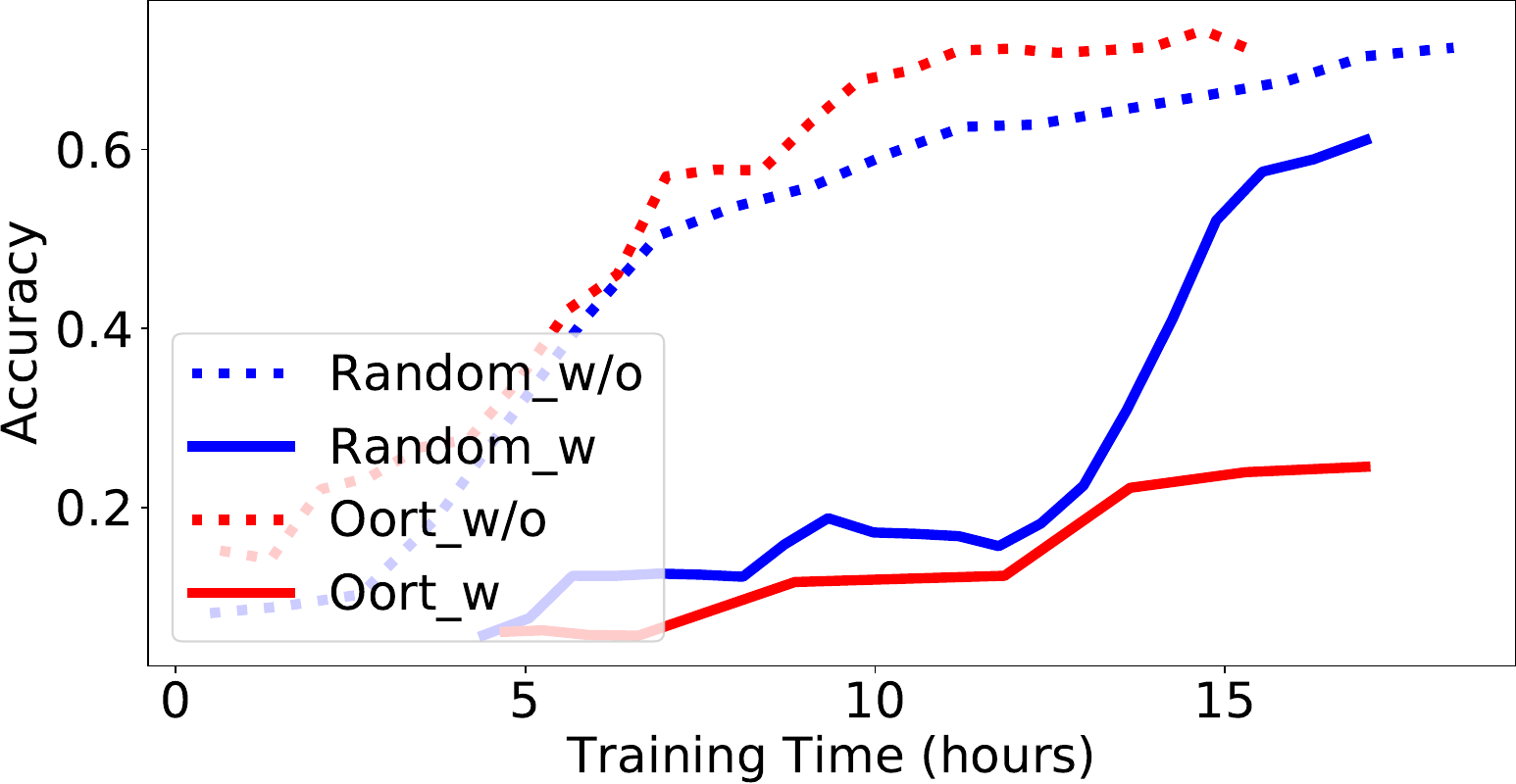}\label{fig:femnist_time_acc_wo_w}}
~
\subfigure[Prediction results with different window sizes in LSTM model: larger window size yields better prediction. ]{\includegraphics[width=0.42\linewidth]{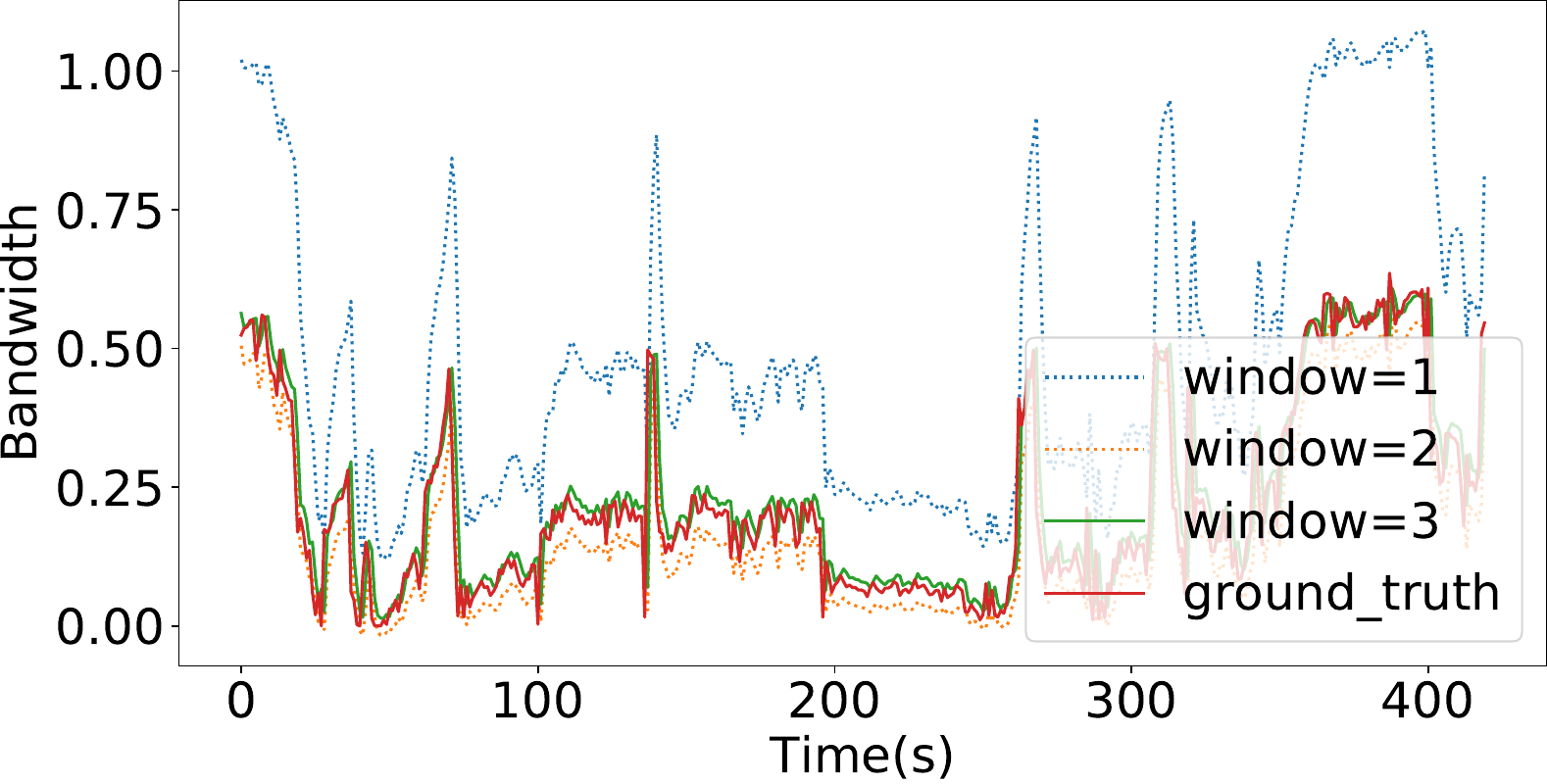}\label{fig:single_step_prediction}}
\caption{Illustration of the limitation of SOTA work and challenge of designing \ours.}
\vspace{-15pt}
\label{fig:fig1}
\centering
\end{figure}

We conduct an experiment on how real-world bandwidth data~\cite{riiser2012dataset,mei2020realtime} affects the SOTA FL system~\cite{lai_oort_2021}, with results shown in Figure~\ref{fig:femnist_time_acc_wo_w}. Following the experimental settings in Oort~\cite{lai_oort_2021}, we select 100 clients in each round for the image classification task using the FEMNIST dataset. Our findings indicate that the time-to-accuracy can be extended by 20\%-30\% compared to scenarios without realistic network fluctuations. 
This degradation leads to prolonged delays during training rounds, which occur frequently throughout the training process. 
Such delays could hinder the parameter updates on devices that hold essential data, further exacerbating the difficulties in achieving efficient FL training.


\noindent
\textbf{Overview of Our Approach.}
To improve the FL system efficiency, we propose \ours, an  FL framework designed to actively
manipulate the selection of 
well-performed clients in a real-world \emph{dynamic} network environment. Our approach first leverages a bandwidth prediction model 
to help with client selection, enabling the server to prioritize clients with fast network connections 
and reducing the selection of bottleneck clients. In contrast, existing greedy selection algorithms~\cite{lai_oort_2021,PyramidFL_MobiCom22} are susceptible to dynamic networks in the wild. These algorithms rely on data from a single previous round, making accurate bandwidth prediction impossible, as illustrated in Figure~\ref{fig:single_step_prediction}.  Additionally, the performance of existing greedy algorithms in client selection is compromised by dynamic network conditions, as a client can perform worse due to sudden changes in the network in the next training round.  To address these challenges, we introduce a long-term observation window to gather data to enable accurate bandwidth prediction in \ours. Within this window, we pause the client selection, allowing us to monitor the overall performance of clients and prioritize those with stable network connections and high-quality data. To strike a balance between long-term scheduling  and fine-grain control of client selection, we adaptively adjust the length of the observation window throughout  the training process, optimizing the overall efficiency of the FL system from a long-term perspective. In the end, \ours can work seamlessly with other optimization methods for client selection.
%
We overcome the following challenges in designing \ours:
\begin{itemize}[leftmargin=*]
    \item \textbf{Challenge~\#1 [Accounting for bandwidth]:} It is hard to achieve accurate bandwidth prediction relying solely on data from the previous round in the SOTA solutions. 
    \item \textbf{Challenge~\#2 [Client selection accuracy in a long-term]:} Current greedy strategy in client selection is more susceptible to a dynamic bandwidth environment. 
    \item \textbf{Challenge~\#3 [Trade-off between long-term schedule and  time efficiency]:} There is a trade-off 
    between the confidence of bandwidth prediction and observation time used to make client selection. 
\end{itemize}

We implement  \ours and conduct experiments on four datasets with real-world user data. Compared with the SOTA methods,    
\ours significantly improves the time-to-accuracy in the FL training process.
\ours achieves a higher final test accuracy while preserving user privacy. We implement \ours on top of two FL paradigms with different optimization methods~\cite{lai_oort_2021,lai2021FedScale}, highlighting its compatibility with existing FL approaches to enhance their time-to-accuracy performance in real-world scenarios with dynamic network conditions.
We summarize our contributions as follows:
\begin{itemize}[leftmargin=*]
    \item To overcome Challenge~\#1, we propose a framework for FL in real-world networks with a bandwidth prediction module, which prioritizes clients with better networking conditions. 
    \item To overcome Challenge~\#2, we propose a long-term greedy strategy in each observation window, which temporally freezes client selection, improves the system robustness in a dynamic network, and optimizes the system efficiency with client selection.
    \item To overcome Challenge~\#3, we propose an algorithm to determine the size of the bandwidth observation and data accumulation window.  This algorithm effectively handles the trade-off between client performance evaluation and  client manipulation granularity.
    \item We implement \ours and integrate it with other client selection algorithms, 
    allowing the existing FL paradigms to perform better in real-world dynamic networks. 
\end{itemize}

%% file: 2_background.tex
There are various approaches for optimizing time-to-accuracy while achieving reasonable model accuracy.
 A recent SOTA FL system, called Oort~\cite{lai_oort_2021}, proposes a client selection strategy that considers both data and system heterogeneity in its system design.  The main idea of client selection is to choose clients who have training data that can contribute the most to the model training and finish local training in a short time period. 
Although those methods can improve system efficiency and achieve better time-to-accuracy performance, they assume all the clients upload updates in a stable network. 
\textit{In this work, we show that dynamic bandwidth can downgrade the performance and undermine the optimization methods on client selection.}

We conduct preliminary experiments on Oort
to show how the dynamic network undermines its client selection optimization methods, as shown in Figure~\ref{fig:femnist_time_acc_wo_w}. We use \mbox{FEMNIST}\cite{cohen2017emnist} dataset
to check the time-to-accuracy and final accuracy on the real-world bandwidth data~\cite{hsdpa,nyc}. Our findings are summarized in the following two aspects.

\noindent\textbf{Impractical Client Selection Setting in Wireless Network.}
From the result, we find that the time-to-accuracy in a dynamic bandwidth environment is much longer compared with the previous static-bandwidth data~\cite{lai_oort_2021}. 
This problem motivates us to  utilize a bandwidth prediction model for selecting clients. Without relying on any additional data, collected bandwidth data can be used in guiding client selection. The bandwidth prediction model helps to select clients with high-quality network conditions
and reduces possible bottleneck clients.

\noindent\textbf{Inherent Limitation of Short-term Greedy Algorithm.}
The short-term greedy algorithm in previous work is more susceptible to bandwidth dynamics in a dynamic network environment.
The major limitation of the greedy client selection algorithm is the fact that the clients' model update uploading time in the last round does not indicate the communication quality in future rounds. With dynamic bandwidth data, the client with the highest utility in the last round might become the bottleneck in the future. Here, we use a long-term greedy strategy to collect more data and make a longer observation of client performance. 


%% file: 4_design-detail.tex
\subsection{\ours Overview} 

\begin{figure}[t]
\centering 
\subfigure{\includegraphics[width=0.82\linewidth]{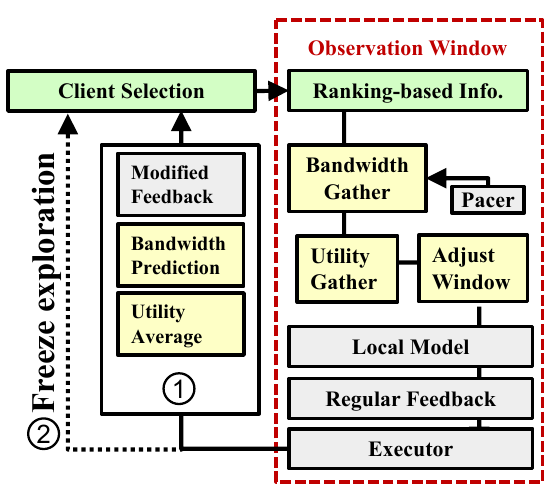}\label{fig:time_acc_wo}}
\vspace{-1pt}
\caption{\ours Framework.}
\vspace{-15pt}
\label{fig:framework}
\centering
\end{figure}


We first provide an overview of how \ours improves the efficiency between chunks of the window and among the training rounds inside the window. Then, we discuss each step and component of \ours in details.

Figure~\ref{fig:framework} shows the \ours architecture with the proposed observation window. \ours aims at gathering data within the observation window and using this data to control client selection based on client utility and bandwidth prediction. Inside the window, \ours accumulates the bandwidth data and the utility score of clients' performance in each round and temporarily freezes the client selection in \textcircled{2}. After several rounds of observation, \ours predicts the bandwidth for each client based on historical bandwidth. The accumulated utility (i.e., the performance of a client) is also averaged to show whether the client has stable and high performance.
The average utility and bandwidth prediction results in \textcircled{1} are used for modifying feedback and selecting clients for the next window. Next, we describe each step of \ours workflow in detail.

The server first initializes the model and dispatches it to each selected client  in the pool, where clients have different configurations~\cite{lai_oort_2021,PyramidFL_MobiCom22,lai2021FedScale}. Then, the server starts watching the performance of each client in the observation window. On the client side, each client starts training the model with local data, followed by transmitting the model updates to the server. All model updates collected by the server are averaged and contributed to the shared model. Based on the training time duration and importance of the update, a utility score (i.e., the ranking-based info) will be calculated and assigned to each device client. The utility score and training duration are accumulated in the observation window. Clients continue to train and update the model and share the updated model in each round but without further client selection. When the number of rounds of training reaches the window size, the server makes predictions over the following rounds based on the bandwidth of each device. The server will select the client based on the history of utility performance and bandwidth prediction results.

\subsection{Prediction of Dynamic Network Conditions}
\vspace{-1mm}

To improve time-to-accuracy, \ours should be able to give rewards to clients who will have a fast network connection speed and penalize clients who might suffer from bottlenecks.
This is not trivial since the server can know a client has become the bottleneck only after a delay has occurred, \textit{how can a server impose a penalty to a client that may become a bottleneck before it appears in the training round?}


To address the challenge, 
we propose to establish a communication protocol between the server and clients considering the bandwidth prediction method. 

\subsubsection{Bandwidth prediction module} 
Our first challenge is to obtain the bandwidth data without requesting any additional information
in order to preserve privacy.


\noindent
\textbf{Bandwidth Calculation.} As the server has the knowledge of the time cost of each client, we can calculate the bandwidth information as follows:
\vspace{-1mm}
\begin{equation}
\begin{aligned}
    T\left(C_{i}, R_{i}\right)=\Delta \operatorname{Comp}\left(C_{i}, R_{i}\right)+\Delta \operatorname{Comm}\left(C_{i}, R_{i}\right) \\
    \Delta \operatorname{Comm}\left(C_{i}, R_{i}\right) = \frac{\mathbb{U}(pull)+\mathbb{U}(push)}{b_{t}}.
    \label{bandwidth_cal}
\end{aligned}
\end{equation}
\vspace{-1mm}

Here, the time $T\left(C_{i}, R_{i}\right)$ that each client $C_{i}$ consumes in round $R_{i}$ comprised of two components: server-client communication $\Delta \operatorname{Comm}$ and local client computation $\Delta \operatorname{Comp}$. Previous work~\cite{PyramidFL_MobiCom22} showed that the communication step usually takes much more time (about 90\%) than device computation (about 10\%). For each client, communication time is proportional to the bandwidth, because the model update size $\mathbb{U}$ that a client needs to upload changes slightly during the training.
Thus, we directly use the time cost of each client in the bandwidth prediction instead of collecting more data.

\vspace{1mm}
\noindent
\textbf{Bandwidth Penalty.}
In \ours, after collecting clients' data for several rounds of training, the server can use the historical data for drawing predictions. The prediction will be used in modifying feedback, the utility score for client $i$, for client selection in the following round. Based on the existing feedback function in Oort, we include an additional factor in the feedback as follows:
\begin{equation}
\begin{aligned}
\operatorname{Util}(i)=\underbrace{\left|B_{i}*\mathbb{F} \right| \sqrt{\frac{1}{\left|B_{i}\right|} \sum_{k \in B_{i}} \operatorname{L}(k)^{2}}}_{\text {Statistical utility U(i)}} \times \underbrace{\left(\frac{T*\mathbb{F}}{t_{i}}\right)^{\mathbb{1}\left(T<t_{i}\right) \times \alpha}}_{\text {System utility }}
\label{bandwidth_feedback}
\end{aligned}
\end{equation}

\begin{equation}
    \mathbb{F}= Norm(P(b_{H}))
    \label{bandwidth_norm}
\end{equation}
\vspace{-1mm}
In the statistical utility part, $B_{i}$ is the local data samples set, and $\operatorname{L}(k)^{2}$ is the training loss of data sample k. We add the bandwidth prediction result $\mathbb{F}$  as a factor to adjust statistical utility in feedback.
In the system utility part, $T$ is the developer-preferred duration, $t_{i}$ is the wall clock time duration for  client $i$ and  $\alpha$ is the penalty factor. Oort uses  $\mathbb{1}(x)$ as an indicator function that takes value one if $x$ is true and 0 otherwise. We use $\mathbb{F}$ to adjust the system utility, where $\mathbb{F}$ is the normalized result of bandwidth prediction model $P$ on data history $b_{H}$.

In Eq.~(\ref{bandwidth_feedback}), the bandwidth is important in selecting clients for the next round. The result of bandwidth prediction is normalized due to the different ranges of bandwidth on different devices.
We prefer to select clients in a fast network with high utility. 
Our reward and penalty factor allows us to exclude a client from training when its network quality is inadequate.

\noindent
\textbf{Offline Bandwidth Prediction Model.}
\ours uses an LSTM model to predict the bandwidth of clients in the next round. 
To ensure that the offline model meets the memory and latency requirements, we use the lightweight three-layers LSTM model. Given hundreds of bandwidth traces in the FL training process, to make \ours privacy-preserving, we only use one bandwidth trace to train the offline model and leave the other hundreds of client bandwidth traces available for clients in the system evaluation. More details on the mechanism of assigning bandwidth traces to devices can be found in Section~\ref{sec:evaluation}.

\input{alg}
\subsubsection{Bandwidth prediction in \ours}
Algorithm~\ref{alg:1} shows how the bandwidth prediction module 
cooperates with the current client selection strategy in making decisions in a dynamic network environment. 
Since Oort~\cite{lai_oort_2021} selects clients based on client utility and time duration, our goal is to modify the client duration time with our bandwidth prediction module~(Lines 4-26).
Inside the prediction function, \ours uses an observation window to keep the bandwidth data for every client at the current time, which comes from the communication data used in the current round~(Lines 9-10). After collecting all the bandwidth data of the last round, the model continues to update the parameters and train for the next round but stops selecting new clients~(Lines 8-12). The server starts to make client selections until the number of observed rounds is equal to the window size. It uses the bandwidth history to predict the communication of each client for the next round~(Lines 16-25). After predicting the bandwidth, we compute a reward or penalty that is dependent on the ratio between the predicted bandwidth and a predefined threshold.
This reward or penalty is then incorporated into the training duration time and the client utility calculation.

When applying the reward and penalty scores, $\alpha$ increases as a reward when the bandwidth prediction result is higher than the reward threshold. Thus, the server increases the priority of those clients~(Line 20). The penalty score reduces the occurrence of bottleneck clients when the prediction result is close to zero~(Line 23). When the prediction result locates between the award/penalty thresholds, \ours does not modify the client feedback~(Line 25).
While the SOTA solutions use exploration and exploitation strategy in client selection to keep clients with high utility scores and explore for new clients, \ours utilizes varying award and penalty scores, allowing some devices to remain in the training due to the high client utility and future communication ability.

\begin{algorithm}
\DontPrintSemicolon
  \KwInput{Client set~$\mathbb{C}$, local training iteration base \textit{i}, bandwidth penalty threshold $TH_{L}$, bandwidth reward threshold $TH_{H}$, current round $R_{i}$, bandwidth prediction model \textit{P}, client utility set $\mathbb{U}$, client time duration set $\mathbb{D}$, obseavation window size W, adjustment coefficient \textit{c}.}
  \KwOutput{Modified Feedback ($\mathbb{U}$, $\mathbb{D}$).}
  \tcc{Initialize global variables}
  $ {B}_{H} \gets $ \textbf{Function} Observation($\mathbb{C}$,$R_{i}$,W) \\ 
  $ \mathbb{U},\mathbb{D} \gets $ PerformanceFeedback($R_{i}$)\\
  $\mathbb{U}, \mathbb{D} \gets $ \textbf{Function} Prediction(${B}_{H},\mathbb{U},\mathbb{D}$)\\ 
  $\mathbb{C}_U \gets ClientSelection~(\mathbb{U}, \mathbb{D}, \mathbb{C})$\\
\textbf{Function} Observation ($\mathbb{C}$,$R_{i}$,W)\\
\tcc{Initialize bandwidth historical set}
\Indp
{
    $B_{H} \gets \emptyset$ \\
    \While ($R_{i} \mod W$)
    {
      \tcc{Record bandwidth for each client}
    \Loop ( client $j \in \mathbb{C}:$){
        $t \gets \text{CurrentTime()}$ \\
        $b^{j}_{t} = \text{GetCommData}(j, t)$ \\
        $B^{j}_{H}.\text{append}(b^{j}_{t})$ \\
    }
    $i \gets i+1$ \\
    Freeze Client Selection \\
    \Comment{Continue Model Training and Model Update}
    }
    \textbf{Return} $B_{H}$ \\
}
\Indm
\textbf{Function} Prediction ($\mathbb{C}$,$TH_{H} $,$TH_{L}$,$R_{i}$,\textit{P},$\mathbb{U}$,$\mathbb{D}$,\textit{c})\\
\Indp
     \tcc{Bandwidth prediction for each client}
     \Loop ( client $j \in \mathbb{C}:$)
    {
    $\alpha = P(B^{j}_{H})$ \\
    \eIf{$\alpha \geq TH_{H} $}
    {
        $\alpha \gets (-log(1-\alpha)+c) $
    }{
        \eIf{$\alpha \leq TH_{L} $}
        {
            $\alpha \gets exp(\alpha+c)$; 
        }{
        $\alpha \gets 1 $;
        }
    }
    \tcc{Update client selection feedback}
    $ \mathbb{U}(j) \gets \mathbb{U}(j) \times \alpha$ \\
    $ \mathbb{D}(j) \gets \frac{\mathbb{D}(j)}{\alpha}$  \\
    }
    \textbf{Return} $(\mathbb{U}, \mathbb{D})$ \\
\Indm
\caption{Bandwidth Prediction in FL.}
\label{alg:1}
\end{algorithm}

\vspace{-1mm}
\subsection{Long-Term Scheduling Strategy}
\vspace{-1mm}

In this section, we analyze the long-term scheduling strategy in dynamic network environments and design an algorithm to offer a better client selection method. The clients with the current best performance found by the greedy algorithm cannot guarantee the best selection in the long run. 
The short-term client selection suffers from the instability caused by the dynamic bandwidth data. 
A device that underperforms due to a temporary network fluctuation is less likely to participate in future training. The goal of \ours is to provide a longer observation window to slow down the decision-making process.
The long-term greedy strategy in \ours makes the system more robust and efficient.

In Algorithm~\ref{alg:2}, we show the long-term greedy strategy in \ours. The goal is to evaluate the overall performance of each client who participates in the past several rounds. We use an accumulation window, to sum up the time duration for each client at every round. This average value represents the overall utility of a client. 
We freeze the client selection when accumulating the data. The server resumes client selection  until data is collected enough to make decisions based on the overall understanding of each client.
In this way, we  select clients with stable performance and high utility scores. 

\begin{algorithm}

\DontPrintSemicolon
  \KwInput{Client set~$\mathbb{C}$, local training iteration base \textit{i}, current round $R_{i}$, observation window size W.}
  \KwOutput{General time duration set of client performance $\mathbb{D}$.}

\textbf{Function} LongTermGreedy ($\mathbb{C}$,$R_{i}$,W)\\
\Indp
    
    \While($R_{i} \mod W$)
    {
         \tcc{Keep time cost of every client}
    \Loop(client $j \in \mathbb{C}:$)
    {
            ${d}^{j}_{t} \gets \text{ClientTimeDuration}(j)$ \\
            ${D}^{j} \gets {D}^{j}+ {d}^{j}_{t}$ \\    
    }
    Freeze Client Selection; \\
    $i \gets i+1$\\
    }  
    
    \Loop(client $j \in \mathbb{C}:$)
    {
    ${D}^{j} \gets \frac{D^{j}}{W}$ \\
    }
    \textbf{Return} $D$\\
\Indm

\caption{Long-term Greedy Strategy.}\label{alg:2}
\end{algorithm}

\subsection{Trade-off Between Bandwidth Prediction and Long-Term Schedule}
The efficiency of \ours depends on two factors: the prediction accuracy of bandwidth and the window size of data accumulation in the long-term greedy strategy. However, a trade-off exists on the window size of the server watching data for prediction and accumulating data before making client selection. 
Intuitively, we can improve bandwidth prediction by using a larger window with more bandwidth data. However, the long observation window causes a longer delay for the server to make necessary adjustments in client selection. On the contrary, a short window can force the server to make quick updates on removing lagged clients, but less data collected to predict bandwidth lowers the accuracy of client selection.

Our goal is to optimize the window size for monitoring bandwidth history data and data buffering. We apply an adaptable approach in adjusting the window size, which takes the system performance with the previous window size into account.
We formulate the problem as follows:
\vspace{-1mm}
\begin{equation}
W_{\text{opt}} = \underset{W}{\operatorname{argmax}} \left( P([b_{1}, ..., b_{W}]) \times \frac{c}{\mathbb{T}_{W}} \right)
\end{equation}
Our goal is to find the optimal window size ${W}_{\text {opt }}$ that can well predict the historical bandwidth data $b_{1},...,b_{W}$ with a short observation time $\mathbb{T_{W}}$.
\vspace{-1mm}

Next, we present Algorithm~\ref{alg:3} to optimize the window size in \ours. 
We aim to adjust the size of the frozen window based on the overall client performance, which is evaluated by the time duration of the round. We increase the control strength in selecting clients at a faster pace when the global training time is longer than our pre-defined threshold. 
So the server should make a quick update to remove the slow clients and try to find some new clients. When the general network speed becomes faster, we use a longer window in bandwidth observation and data accumulation to obtain better bandwidth prediction and system performance. 

\begin{algorithm}
\DontPrintSemicolon
  \KwInput{Client set~$\mathbb{C}$, local training iteration base \textit{i}, current round $R_{i}$, observation window size W.}
  \KwOutput{Adjusted window size W.}

\textbf{Function} WindowAdjustment($\mathbb{C}$,$R_{i}$,W)\\
\Indp
    $d_{i} \gets \text{GlobalTimeDuration}(i)$ \\
    \eIf{$ {d}_{i} \geq D_{H} $}
    {
        $ W \gets W \times \frac{D_{H}}{{d}_{i}} $
    }{
        \If{$ {d}_{i} \leq D_{S} $}
        {
            $ W \gets W \times \frac{D_{S}}{{d}_{i}}$; 
        }
    }
    \textbf{Return} W\\
\Indm
\caption{Trade-off on Window Size.}\label{alg:3}
\end{algorithm}

By adaptively adjusting the window used for bandwidth prediction and long-term greedy strategy, \ours achieves a local optimum for different network environments in the training process and improves the overall time-to-accuracy.

%% file: 5_evaluation.tex
\subsection{Experimental Setup}
\vspace{1mm}
\noindent
\textbf{Implementation.} 
We develop \ours on FedScale~\cite{lai_oort_2021}, a benchmark framework for FL based on Oort\cite{lai_oort_2021}. FedScale can simulate the computation and communication over different types of devices. Although FedScale uses real computation capacities and network connection data, they are fixed through the training process without considering the network dynamics. We introduce the bandwidth dynamics into FedScale. 
We use Yogi~\cite{reddi2021adaptive} as the  FL aggregation method and build our model with PyTorch on top of FedScale. We use the lightweight LSTM model to help predict bandwidth using hidden size as 2 and a learning rate of 0.01.

\begin{figure}[t]
\centering 
\subfigure[Cumulative Distribution Function for various trace data.]{\includegraphics[height=1.1in]{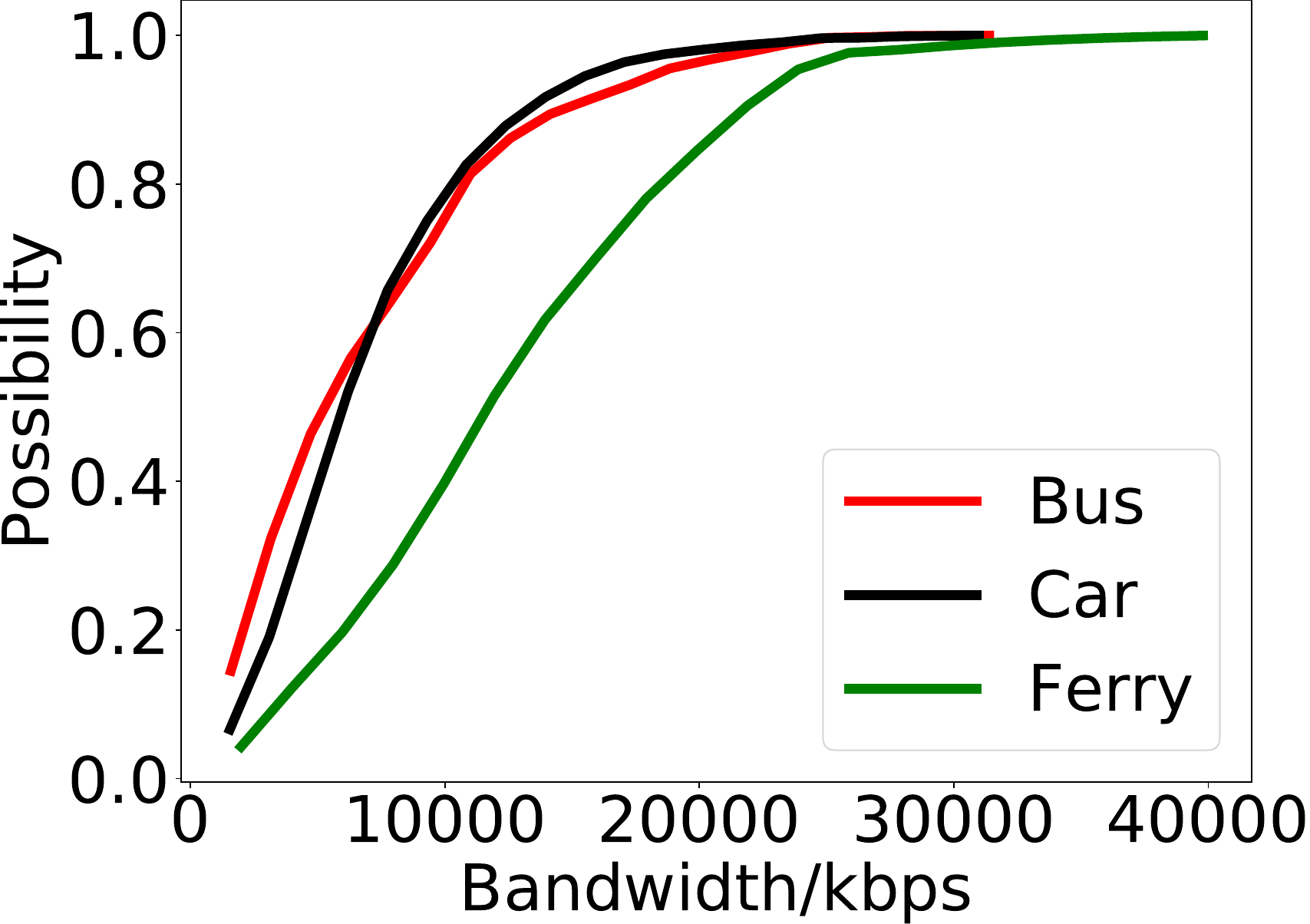}\label{fig:cdf}}
~
\subfigure[LSTM model prediction loss on test dataset with different window sizes.]{\includegraphics[height=1.1in]{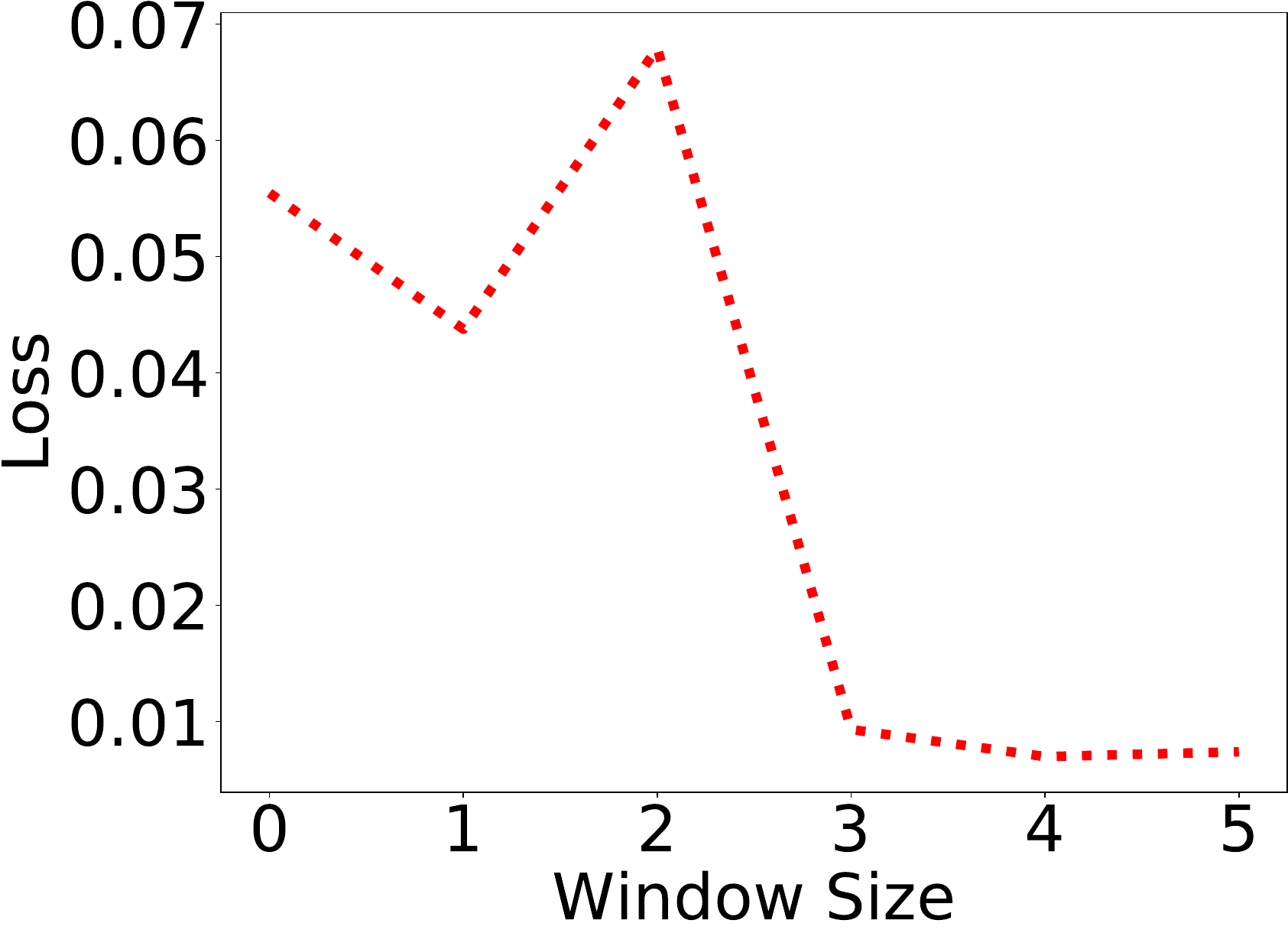}\label{fig:lsmperdiction}}
\caption{Raw trace data CDF for selected transportation types and LSTM prediction loss.}
\vspace{-15pt}
\label{fig:fig3}
\centering
\end{figure}

\vspace{1mm}
\noindent
\textbf{Datasets.}
To simulate the real-world network activity, we extract bandwidth data from two classes of traces: the HSDPA dataset~\cite{riiser2012dataset} and the NYC dataset~\cite{mei2020realtime}. These two datasets are large real-time bandwidth datasets, including transportation traces on the train, ferry, car, bus, and metro. For each data trace, it records the bandwidth between an LTE mobile phone and a remote server every second. We show the raw data distribution in Figure~\ref{fig:cdf}. There are multiple experiments at different times of the day to improve the credibility of the dataset. In the evaluation, we map each client in the client pool with a device bandwidth trace from the two bandwidth datasets using the division method of hashing. Therefore each client can upload their model updates with real-time high-fluctuated bandwidth.



To evaluate \ours, we use four datasets with various real-world trace data and client configurations:
\begin{itemize}[leftmargin=*]
    \item \textit{Image Classification}~\cite{kuznetsova2018open,cohen2017emnist}. For the two computer vision tasks, we evaluate \ours on two image datasets: OpenImage~\cite{kuznetsova2018open} and FEMNIST~\cite{cohen2017emnist}. OpenImage dataset has 8,000 clients collaborating on 1.1 million images. We use the MobileNet~\cite{kuznetsova2018open} as the training model with Top-1 accuracy for image classification. The FEMNIST dataset has 3,400 clients and contains 640K images, in which we also use the Top-1 accuracy.
    \item \textit{Speech Recognition}~\cite{warden2018speech}. In this task, We evaluate \ours on the Google Speech recognition task~\cite{warden2018speech}. We conduct a training experiment with 2,618 clients and 105K audio commands. We use ResNet-18 ~\cite{he2016deep} for our 20-class speech recognition task.
    \item \textit{Human Activity Recognition}~\cite{ouyang2021clusterfl}. We use the HARBox dataset~\cite{ouyang2021clusterfl},  containing 34,115 data samples collected from 121 smartphones for human activity recognition, with a two-layer DNN model in the task.
\end{itemize}

\begin{table*}[!t]
\begin{center}
\caption{Summary of \ours's improvement on time-to-accuracy over \sota~\cite{lai_oort_2021}. We compare the overall improvement in final accuracy and  wall clock time to reach the final accuracy.}
\begin{tabular}{ccc|cc>{\columncolor{yellow!30}}c>{\columncolor{yellow!30}}c}
\hline
Federated &\multirow{2}{*}{Dataset}&\multirow{2}{*}{Model}&\multicolumn{2}{c}{\sota~\cite{lai_oort_2021}+Yogi~\cite{reddi2021adaptive}} &\multicolumn{2}{c}{\ours+Yogi}\\
Applications &&& Acc& Time& $\Delta$Metric & TimeCost \\
\hline 
Image Classification & \#1~\cite{kuznetsova2018open} & MobileNet\cite{sandler2018mobilenetv2}&35.84\% &33.77h &\textbf{2\%$\uparrow$}&\textbf{76.9\%}
\\\hline 
Speech Recognition & \#2~\cite{warden2018speech} &ResNet-34\cite{he2016deep} &58.71\%&116.48h&\textbf{5\%$\uparrow$} &\textbf{73.5\%}
\\\hline 
Image Recognition & \#3~\cite{cohen2017emnist}  &Shufflenet~\cite{shuffelnet}&74.91\%&145.28h&\textbf{3$\%\uparrow$} &\textbf{16.3\%} 
\\\hline
Activity Recognition & \#4~\cite{ouyang_clusterfl_2021} &Customized&68.9\%&1.76h&\textbf{2\%$\uparrow$} &\textbf{84.1\%}
\\\hline
\end{tabular}
\vspace{-4mm}
\label{tbl:overall-perf}
\end{center}
\end{table*}

\noindent
\textbf{Metrics and Baselines.} 
We follow the same evaluation metrics as in the SOTA FL solutions~\cite{PyramidFL_MobiCom22,lai_oort_2021}, i.e., time-to-accuracy and final model accuracy. The time-to-accuracy evaluates the wall-clock time for the central server model to achieve a certain accuracy. The model accuracy is the accuracy that the model achieves when it converges. We use the top 1 accuracy by default.
We compare \ours  with two baselines: random client selection methods and the Oort~\cite{lai_oort_2021} client selection method. We use the SOTA optimizer Yogi~\cite{reddi2021adaptive} in the training process. 

\noindent
\textbf{LSTM Model.} We train a three-layer LSTM model with only one airline passengers bandwidth data trace and test the model with other trace data. From Figure~\ref{fig:lsmperdiction}, we find the test loss with a window size much higher than that with a window size of 5. We utilize this bandwidth prediction model in \ours to help to better select clients.

\noindent
\textbf{Parameter Configurations.} We follow the same parameter setting  in previous work~\cite{lai_oort_2021,lai2021FedScale,PyramidFL_MobiCom22}. We use the batch size 20 for all the datasets, for both training and testing. \ours selects 100 clients from 130 candidates in each round by default for OpenImage, FEMNIST, and Speech tasks. Due to the limitation of the size of clients in HAR, we select 5 clients for the task~\cite{shin2022fedbalancer}.
Each client needs to train the local model for 20 epochs on each round, and we test the model every 10 rounds. The learning rate for the FEMNIST dataset is 0.01, and for the remaining dataset is 0.005.

\subsection{Speedup Performance}
\begin{figure}[t]
\centering 
\subfigure[FEMNIST+Yogi]{\includegraphics[height=1.2in]{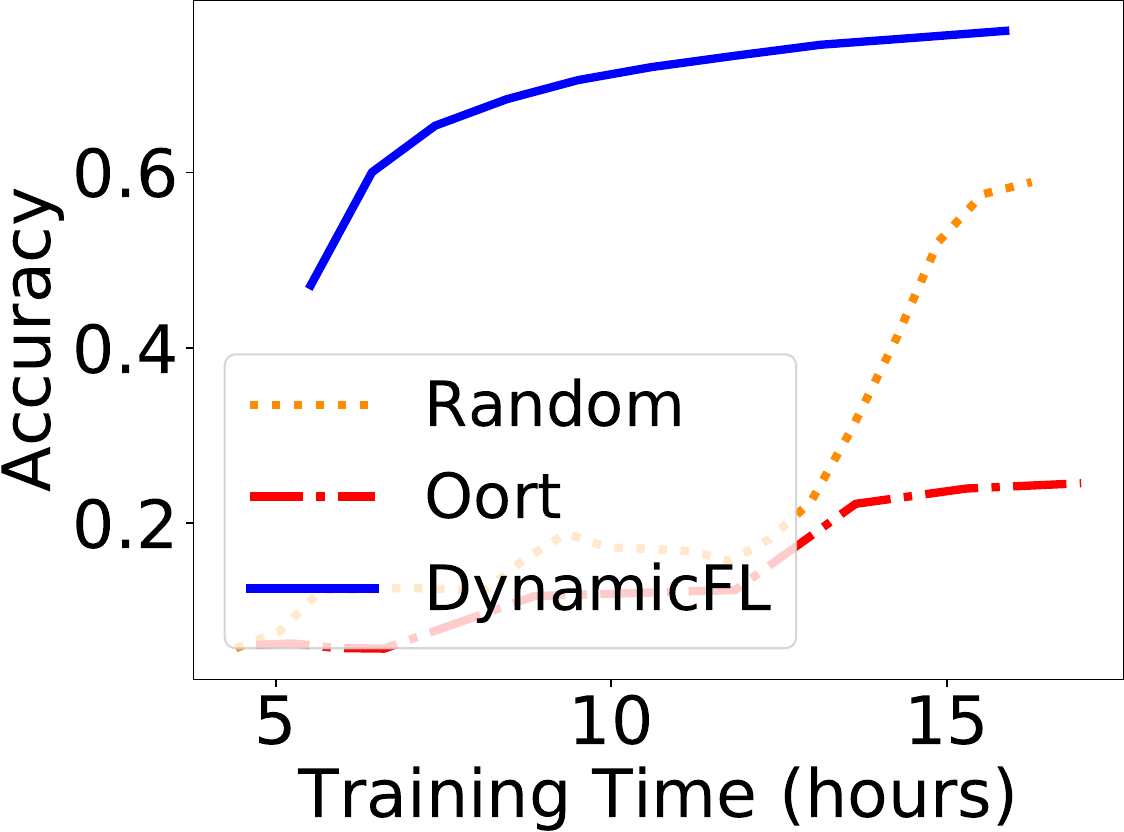}\label{fig:time_acc_wo}}
~
\subfigure[OpenImage+Yogi]{\includegraphics[height=1.2in]{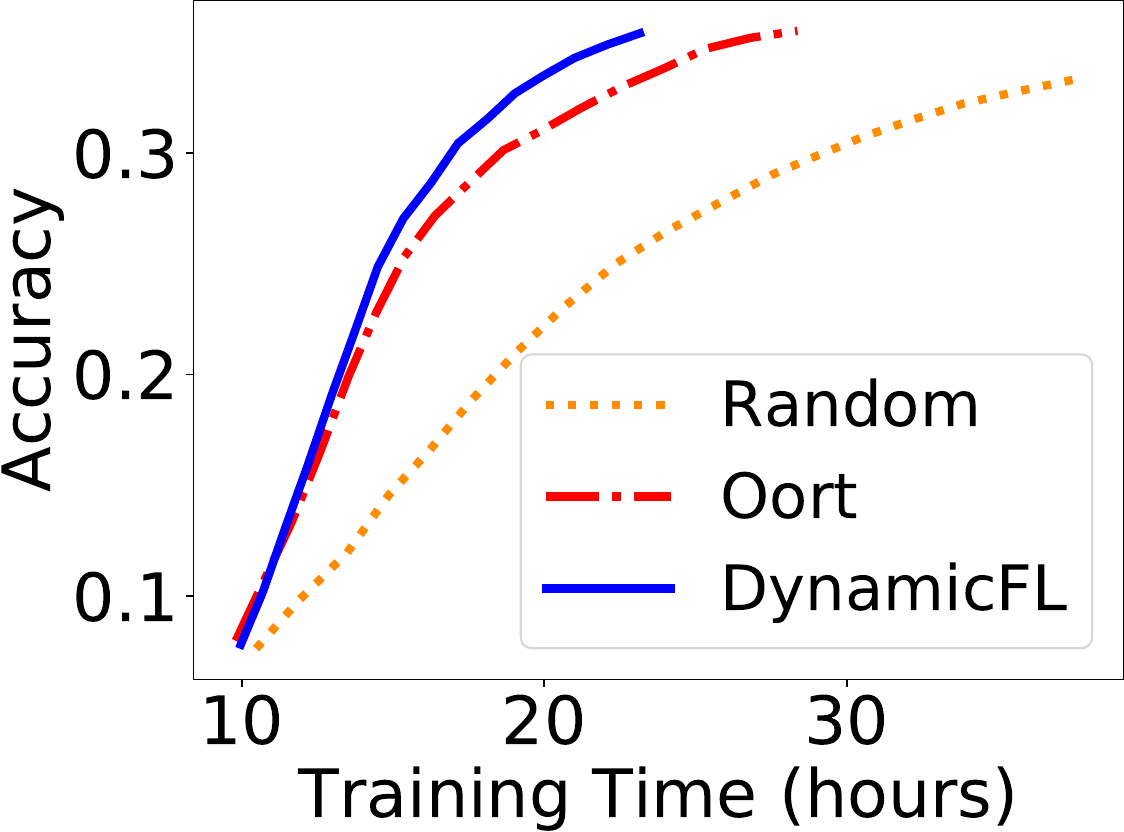}\label{fig:M_S_perp}}
\caption{Time-to-Accuracy for \ours and Oort on FEMNIST and  OpenImage dataset.}
\label{fig:time_acc_openimage}
\vspace{-10pt}
\centering
\end{figure}

\vspace{-1mm}
\noindent
\textbf{\ours Reduces Wall-clock Time to Achieve Final Accuracy.}
We show a clear wall-clock time reduction of \ours with the two baseline models on all four experiments, as shown in Table~\ref{tbl:overall-perf}.
On the OpenImage dataset with real-world device traces and client data, \ours  is 6.13$\times$ faster than the Oort approach, consumes 16.3\% of the wall-time clock time, and improves the final model accuracy. This largest improvement is caused by the diverse data distribution in the OpenImage dataset. The high non-independent and identically distributed data are better exploited with our long-term greedy strategy and bandwidth prediction module. The non-i.i.d. feature of the OpenImage dataset also makes the Oort model susceptible to the dynamic network environment. On the contrary, the time efficiency of \ours on the HAR dataset, with a lower non-i.i.d. level, is the least improved one with only 1.19$\times$ speed-up. When each client shares a small number of samples, the long-term observation data has little difference from the short-term observation data. Figure~\ref{fig:time_acc_openimage} shows more details of the accuracy change along with training time, corroborating the outstanding performance of \ours with dynamic bandwidth data. Compared with \ours, both baseline client selection methods suffer from the greedy selection method with the fast-changing dynamic network, where the client prioritized in the last round can cause a huge lag in the next round of training.
 
\begin{figure}[t]
\centering 
\subfigure[FEMNIST+Yogi]{\includegraphics[height=1.2in]{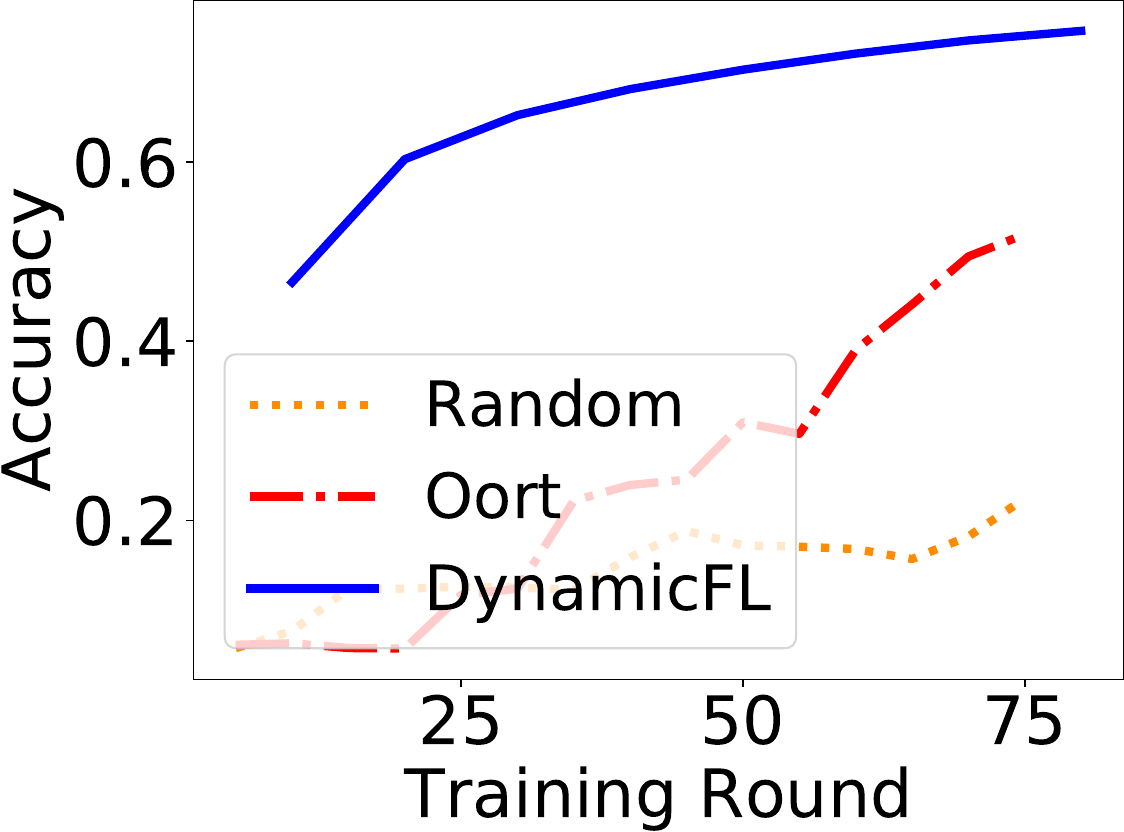}\label{fig:time_acc_wo}}
~
\subfigure[OpenImage+Yogi]{\includegraphics[height=1.2in]{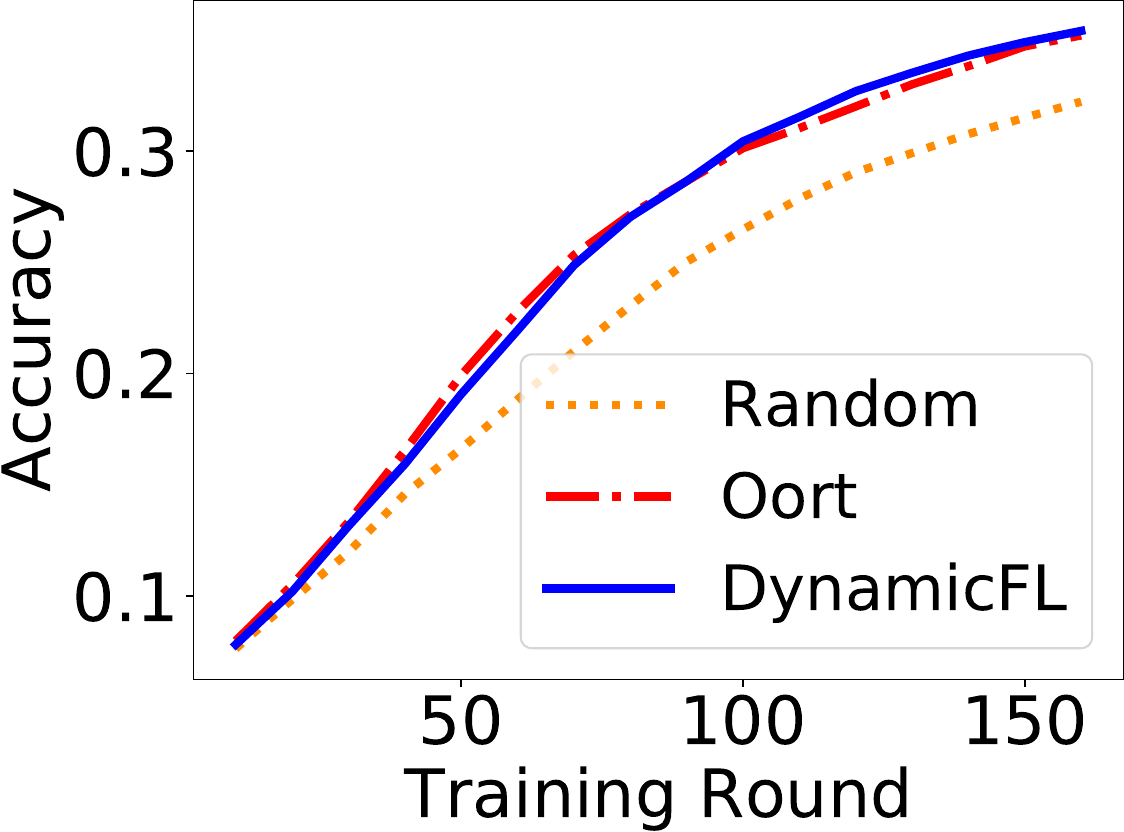}\label{fig:M_S_perp}}
\caption{Round-to-Accuracy for \ours and Oort on FEMNIST and  OpenImage dataset.}
\vspace{-20pt}
\label{fig:round_acc_openimage}
\centering
\end{figure}

\ours reduces the time wasted in waiting for the client with bottleneck network connectivity in the training round and reduces the overall training time. Without the prediction ability on the future bandwidth, the client with a high utility score in the last round can be selected due to the short-term greedy strategy, while causing a delay in the current network environment.
From Figure~\ref{fig:round_acc_openimage}, we can find that \ours does not reduce the number of training rounds significantly to achieve its final accuracy. This can be explained by reducing the time wasted on waiting for the bottleneck clients in each round, as communication time is the major time cost in FL model training.

\noindent
\textbf{\ours Improves the Final Model Accuracy in Limited Training Time.}
From Table~\ref{tbl:overall-perf}, \ours slightly improves the system performance in terms of the final accuracy of the training model. The main reason is previous design gives enough time on waiting for the bottleneck client, so they can finally upload the updates to the server. Therefore, the difference in the final accuracy is limited. Considering the setting in the real-world scenario,  the model accuracy can be improved by 200\% when we stop model training in 24 hours on the FEMNIST dataset. Though the bandwidth prediction model may not select the clients with high data utility temporarily due to current network connectivity, the long-term greedy strategy can 
explore and exploit those clients by freezing client selection and getting a general view of client performance in the long run. With our improved client selection method, our model is likely to converge into the sub-optimal. Our long-term greedy strategy allows the server to make selections more carefully and deliver a faster convergence towards final accuracy.

\subsection{Ablation Study}
We  evaluate the bandwidth prediction module and long-term schedule module respectively. In the ablation study, we deliver our understanding of the importance of  each part in \ours.
\begin{itemize}[leftmargin=*]
    \item \ours w/o Bandwidth Prediction. We first conduct experiments on \ours without using the bandwidth prediction module. \ours freezes client selection inside the window and makes decisions based on the historical performance of clients. But the server does not consider the possible bandwidth changes when selecting clients for future rounds. As such, clients with network connectivity downgrading can become the bottleneck in the future. 

    \item \ours w/o Long Term Greedy strategy. we disable the long-term greedy method and conduct the bandwidth prediction based on data only from the last round. We aim to select clients based on the client utility and training duration from the last round. Without a general consideration of the client performance along with the quick change of network connectivity, the server is likely to select clients causing delays in future training.

\end{itemize}
\noindent
\textbf{\ours Optimizes Client Selection via Bandwidth Prediction in a Dynamic Network.}
Table~\ref{tbl:ablation} shows the evaluation result on  the image classification tasks with only the bandwidth prediction module compared to Oort with a dynamic network. We find the improvements in speedup is limited for these two tasks. The bandwidth prediction part can improve the client selection with the selection of clients with fast network connection speed  after  observation. 
However, relying solely on data from the last round makes the prediction imprecise and insufficient for effectively guiding client selection, where the performance falls short of achieving the level of \ours.

\vspace{1mm}
\noindent
\textbf{\ours Improves the Time Efficiency Using a Long-term Greedy Strategy.}
The long-term greedy strategy provides the main contribution to the improvement of time-to-accuracy, as shown in Table~\ref{tbl:ablation}.
We explain this result through the fact that the long-term greedy strategy can get a better understanding of the client performance.
This can better evaluate the clients in a dynamic environment and guide selection in the next round. We find the long-term greedy strategy can bring a 4$\times$ improvement on wall-clock time. 

By combining the bandwidth prediction and long-term greedy method in \ours, we achieve a more precise prediction result and get a more accurate evaluation score on clients in the dynamic network environment. These two strategies contribute to better client selection outcomes.

\begin{table}[!t]
\begin{center}
\caption{Summary of \ours's improvement on time-to-accuracy over Oort~\cite{lai_oort_2021}: the ablation study on the  improvement in  final accuracy  and  wall clock time.}
\begin{tabular}{c|>{\columncolor{yellow!30}}c>{\columncolor{yellow!30}}c|>{\columncolor{yellow!30}}c>{\columncolor{yellow!30}}c}
\hline
Dataset+Model &\multicolumn{2}{c}{w/o Long-term} &\multicolumn{2}{c}{w/o Prediction}\\
with Yogi&$\Delta$Metric & Speedup&$\Delta$Metric  & Speedup\\
\hline
\#1+MobileNet&\textbf{1\%$\uparrow$}&\textbf{1.04$\times$}&\textbf{1\%$\uparrow$}&\textbf{1.25$\times$} \\
\hline 
\#2+Shufflenet&\textbf{2\%$\uparrow$}&\textbf{1.01$\times$}&\textbf{1\%$\uparrow$}&\textbf{4.72$\times$}
\\\hline
\end{tabular}
\vspace{-15pt}
\label{tbl:ablation}
\end{center}
\end{table}

\subsection{Robustness and Sensitivity Analysis}
\begin{figure}[t]
\centering 
\subfigure[Time-to-Accuracy]{\includegraphics[height=1.1in]{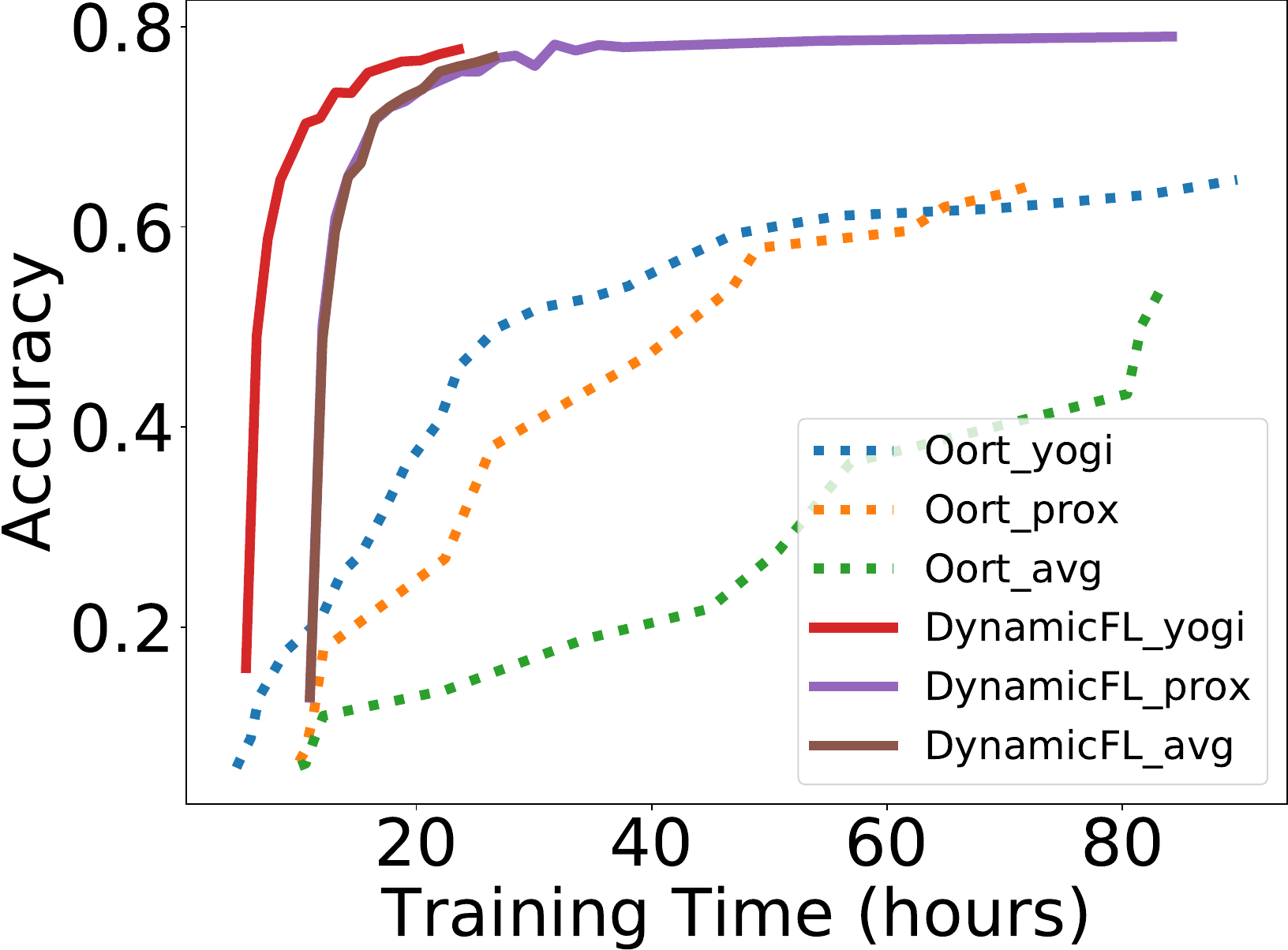}\label{fig:time_acc_wo}}
~
\subfigure[Round-to-Accuracy]{\includegraphics[height=1.1in]{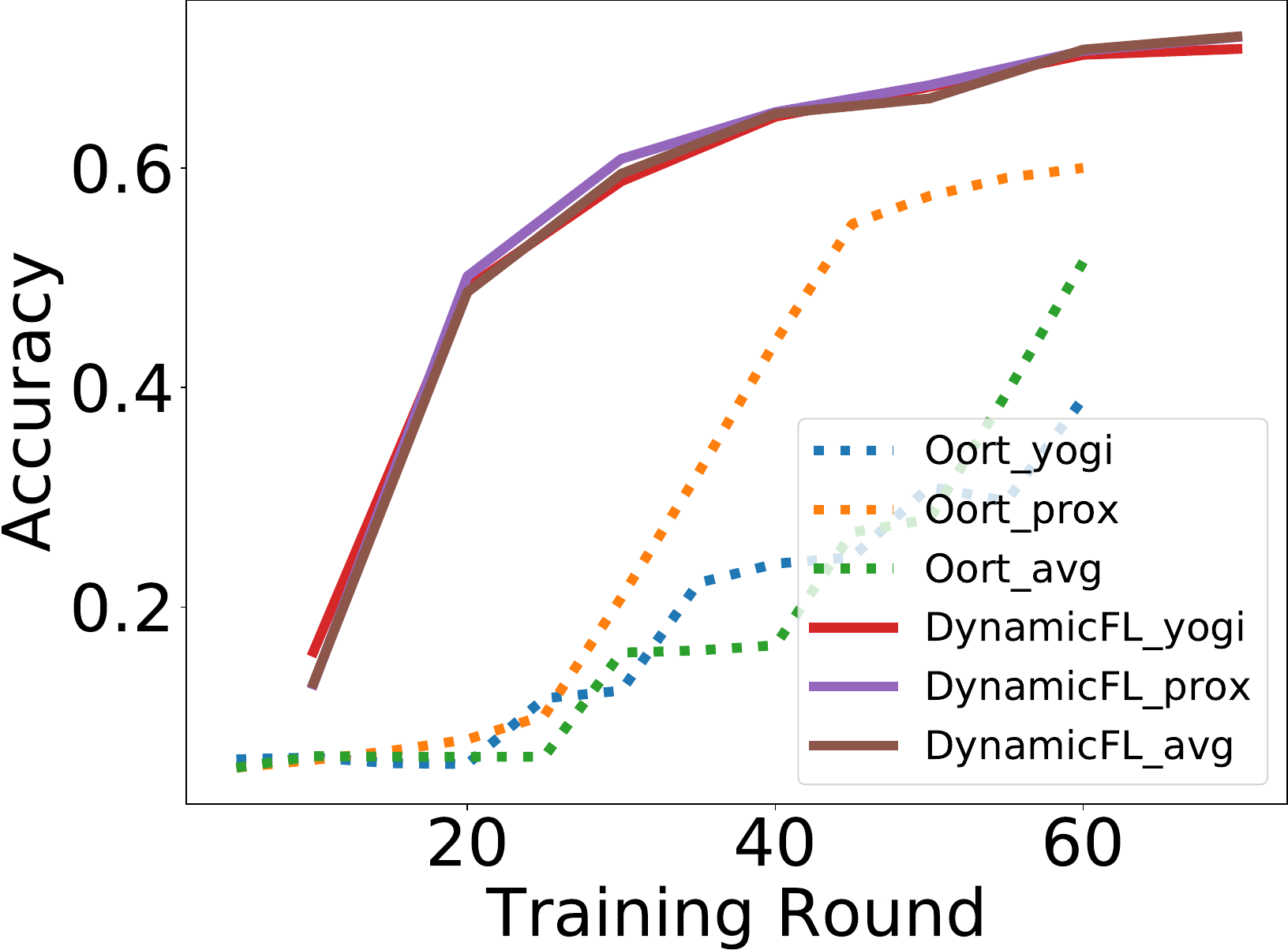}\label{fig:M_S_perp}}
\caption{\ours improves time-to-accuracy with different optimization functions in comparison with Oort.}
\vspace{-10pt}
\label{fig:function_femnist}
\centering
\end{figure}

\vspace{1mm}
\noindent
\textbf{Impact of Different Optimization Functions.}
To examine the impact of different optimizers in our design, we conduct experiments on the FEMNIST dataset~\cite{cohen2017emnist} with three types of optimizers: Prox~\cite{li2020federated}, Yogi~\cite{reddi2021adaptive} and FedAvg~\cite{mcmahan2017communication_fedavg}. From Figure~\ref{fig:function_femnist}, we find that \ours has a better performance on all the three optimizers, compared with Oort\cite{lai_oort_2021} with dynamic bandwidth dataset. The Yogi optimizer is the best one among the three optimizers. We find that the Oort baseline performance is unstable and fluctuates along with the changed network connectivity. With the long-term greedy strategy and bandwidth prediction, \ours removes the bottleneck clients in advance and achieves high convergence with a much shorter wall-clock time.

\begin{figure}[t]
\centering 
\subfigure[Time-to-Accuracy]{\includegraphics[height=1.1in]{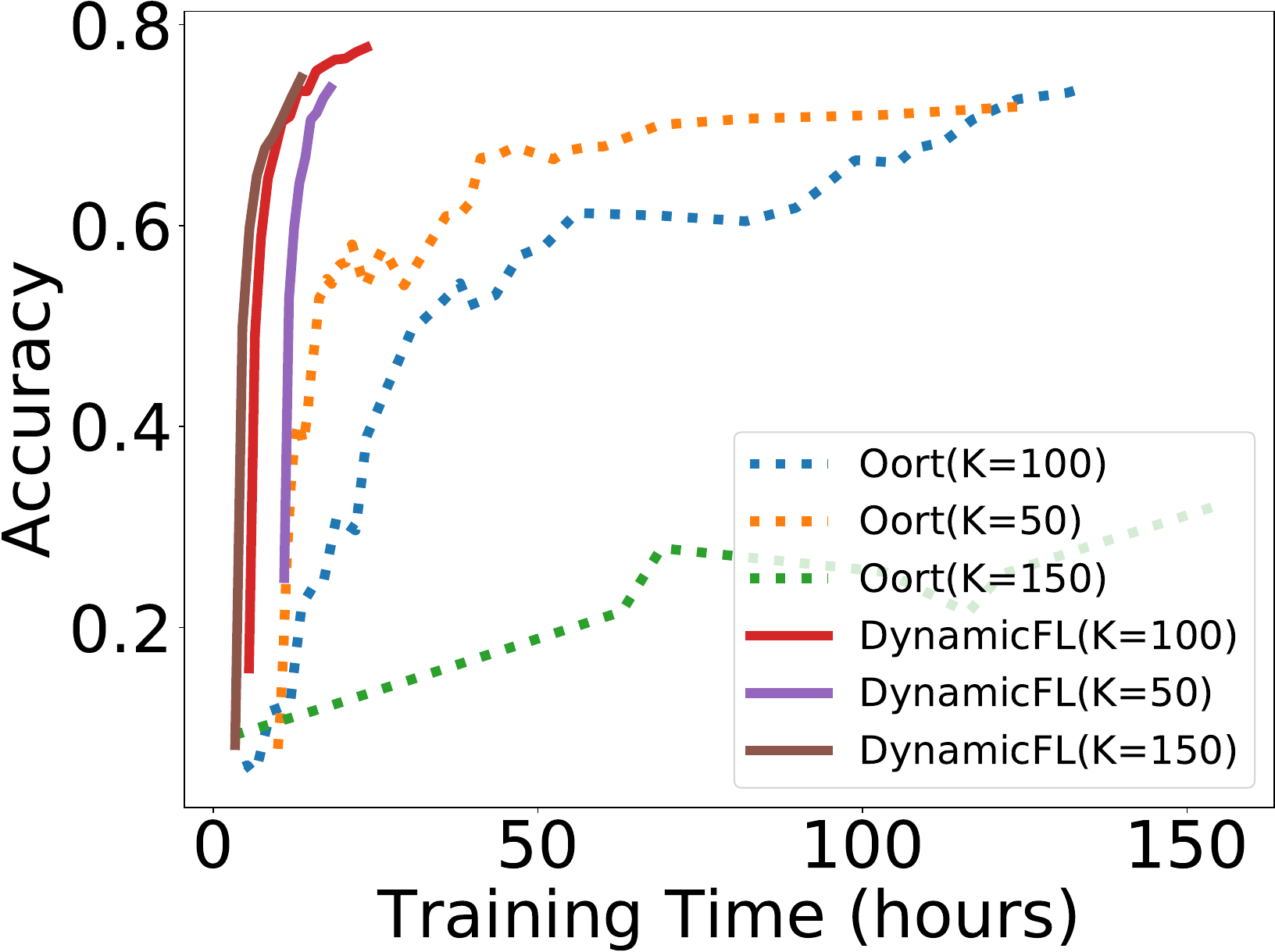}\label{fig:time_acc_wo}}
~
\subfigure[Round-to-Accuracy]{\includegraphics[height=1.1in]{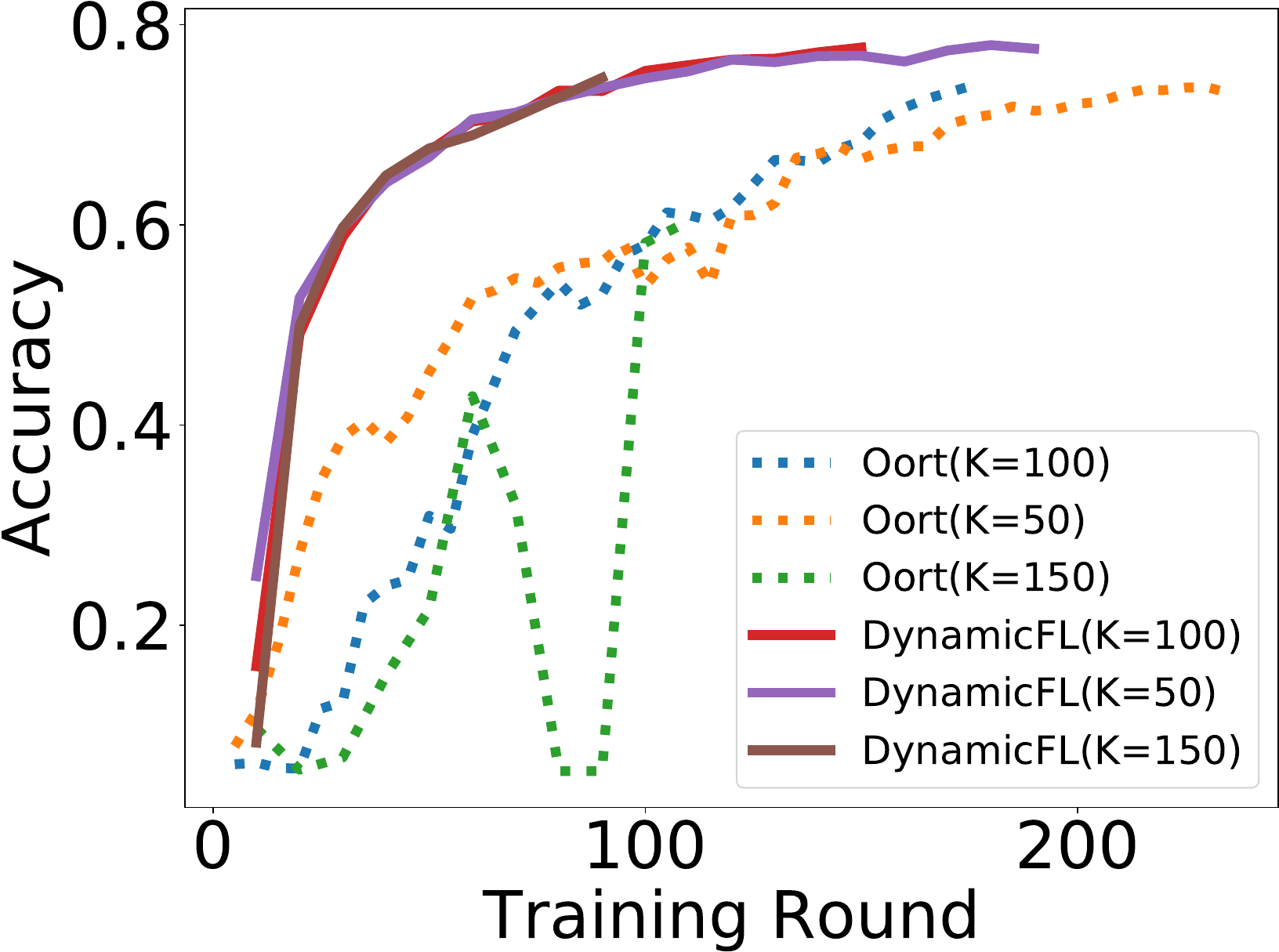}\label{fig:M_S_perp}}
\caption{\ours achieves better performance under the various number of participants.}
\vspace{-15pt}
\label{fig:femnist_client}
\centering
\end{figure}

\noindent
\textbf{Impact of Number of Participants.}
We evaluate \ours with the different numbers of clients participating in the training in each round, viz., 50, 100, and 150. We compare the time-to-accuracy and round-to-accuracy with the Oort model simulated with dynamic bandwidth on the FEMNIST dataset. From Figure~\ref{fig:femnist_client}, we observe that our design can achieve a better model accuracy in far less time while requiring fewer rounds in training with all three settings. Considering the performance with different participant numbers, using more clients leads to greater performance degradation for the SOTA work with real-time bandwidth data. We explain this result through the fact that the increasing number of participants also increases the chance of bottleneck clients appearing in the training round. On the contrary, \ours can remove the slow clients in an effective way
and achieve a similar time-to-accuracy for all three settings.

\begin{figure}[t]
\centering 
\subfigure[Time-to-Accuracy]{\includegraphics[height=1.1in]{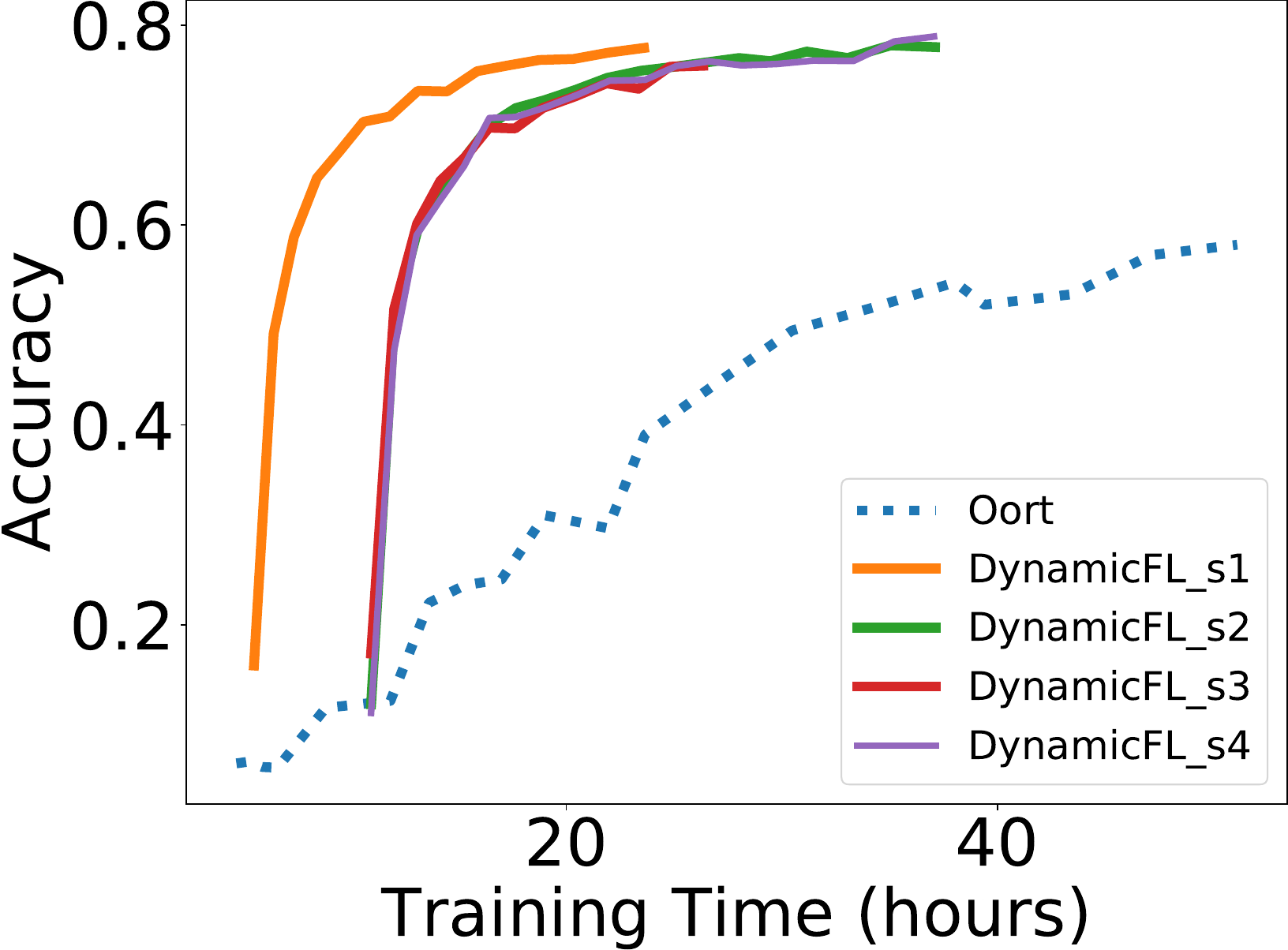}\label{fig:time_acc_wo}}
~
\subfigure[Round-to-Accuracy]{\includegraphics[height=1.1in]{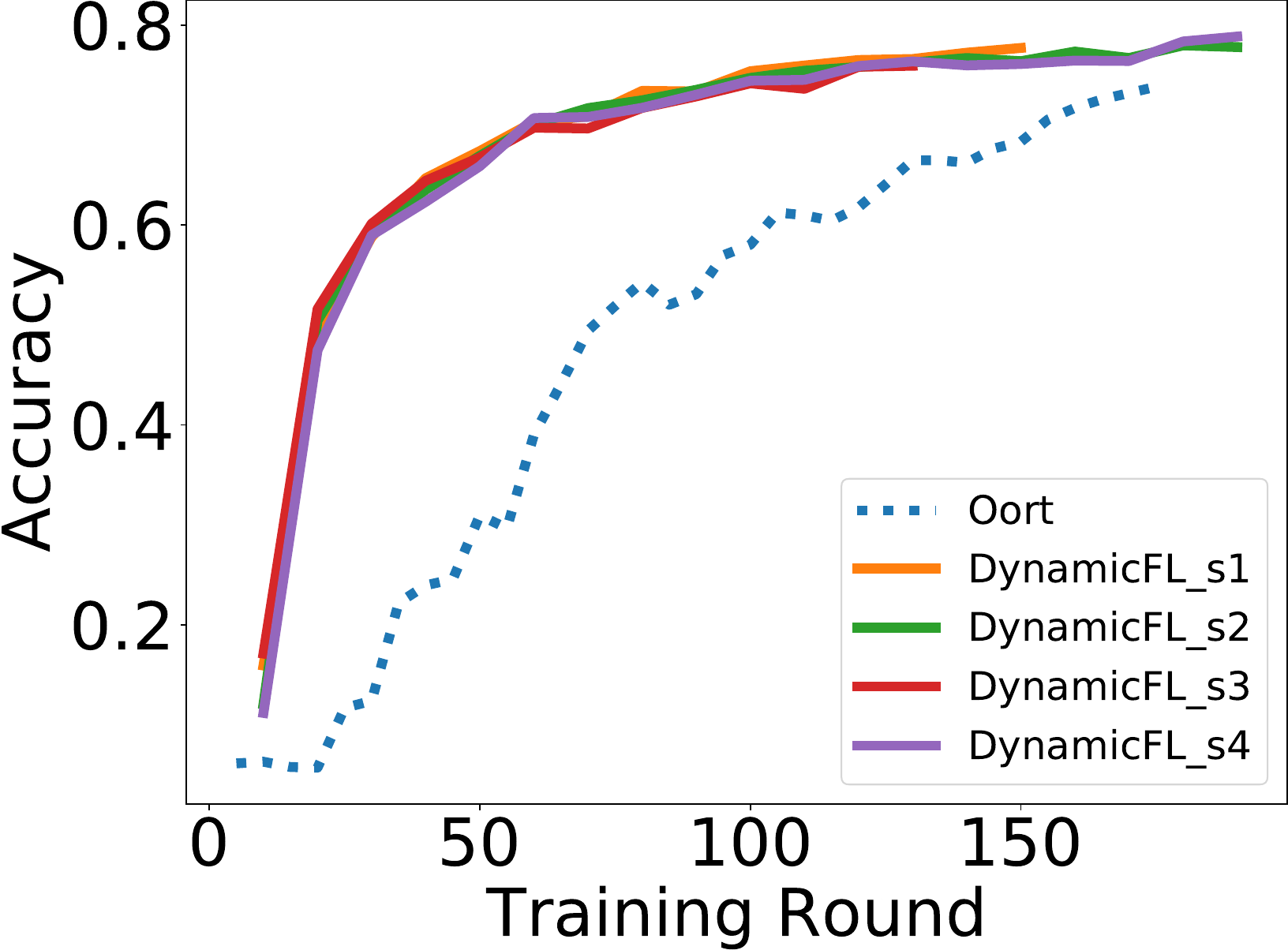}\label{fig:setting}}
\caption{\ours improves the time-to-accuracy across different penalty/reward factors.}
\vspace{-15pt}
\label{fig:femnist_nuber}
\centering
\end{figure}

\noindent
\textbf{Impact of Penalty Score.}
In Algorithm~\ref{alg:1}, \ours uses the reward and penalty factor to update the feedback of clients. We use four settings to check the impact of the coefficient that multiplies the factor of reward and penalty respectively in Figure~\ref{fig:femnist_nuber}. From setting 1 (s1) to setting 4 (s4), the reward and penalty coefficient varies from (1.5,5), (2,6), (2,3), and (1.5,10), where greater reward and penalty coefficients indicate a stronger strength in client manipulation. \ours achieves better performance over Oort on time-to-accuracy with different settings. 
With a selected factor, it can improve performance on the dynamic bandwidth data. 
Although our model does not finish training in fewer rounds, it significantly reduces the time cost in each round.

%% file: 6_relatedwork.tex
\vspace{1mm}
\noindent
\textbf{Bandwidth Prediction in Wireless Networks.}
Bandwidth prediction is an important problem in computer network research, which is first studied as a problem of throughput estimation using historical data~\cite{jiang2002passive, he2005predictability}. 
Most recent studies consider using machine learning models in bandwidth prediction, including the DNN model~\cite{ali2019deep,guo2023black,ivanov2023security,li2021deep}, reinforcement learning model~\cite{mao2017neural},  and LSTM model~\cite{li2021alstm}. In our work, we use an LSTM model in bandwidth prediction to guide client selection in the FL model training. Moreover, we integrate bandwidth prediction with long-term scheduling in client selection to optimize the training efficiency in a dynamic network condition.

\vspace{1mm}
\noindent
\textbf{System Performance Optimization in FL.}
A number of studies have proposed different methods for improving the efficiency of FL systems.
Some studies~\cite{diao2020heterofl,li2021hermes,li2020federated,lai_oort_2021,lai2021FedScale,wang2023vsmask,li2023echoattack,NELoRa_Sensys21} dispatch different models to clients based on their different capabilities to improve the model performance and training efficiency. 
Client selection~\cite{lai_oort_2021} is an emerging approach to optimize both the model accuracy and the time efficiency in FL. 
The follow-up work uses a fine-grained method that considers the importance of different data samples~\cite{shin2022fedbalancer} and adaptive training method~\cite{PyramidFL_MobiCom22} in the synchronized FL system. Although they use real-world capacities for different client devices, none of these studies consider applying FL in real-world dynamic networks. In this research, we are the first to address the problem of client selection with dynamic networks in FL using a real-world bandwidth dataset.

%% file: 7_conclusion.tex
In this paper, we focus on
managing the unstable network
dynamics across massive edge devices in FL by designing a guided client selection strategy.
Its core idea is to combine bandwidth prediction and client selection to improve the training efficiency of FL systems in real-world networks. \ours predicts the network bandwidth of participating clients by observing their training time duration, with which it can coordinate the client selection wisely. 
To optimize the long-term scheduling policy in client selection, we further balance the trade-off between the reliability of network prediction and the client manipulation granularity.
Our evaluation demonstrates \ours's effectiveness under various application scenarios.
Compared with the state-of-the-art FL systems, \ours achieves a better model accuracy while only consuming 18.9\%-84.0\% of the wall-clock time.